\documentclass[fleqn,usenatbib]{mnras}

\usepackage[T1]{fontenc}
\usepackage{orcidlink}

\DeclareRobustCommand{\VAN}[3]{#2}
\let\VANthebibliography\thebibliography
\def\thebibliography{\DeclareRobustCommand{\VAN}[3]{##3}\VANthebibliography}

\usepackage{graphicx}	
\usepackage{amsmath}	
\usepackage{amssymb}	

\newcommand{\frb}{FRB\,20201124A}

\title[FRB\,20201124A host galaxy]{The host galaxy and persistent radio counterpart of FRB\,20201124A}

\author[Ravi et al]{Vikram~Ravi\orcidlink{0000-0002-7252-5485},$^1$ 
Casey J.~Law\orcidlink{0000-0002-4119-9963},$^1$ 
Dongzi Li\orcidlink{0000-0001-7931-0607},$^1$ 
Kshitij Aggarwal\orcidlink{0000-0002-2059-0525},$^{2,3}$ 
Sarah Burke-Spolaor\orcidlink{0000-0003-4052-7838},$^{2,3,4}$ 
\newauthor
Liam Connor,$^1$
T.~Joseph~W.~Lazio,$^{5}$ 
Dana Simard\orcidlink{0000-0002-8873-8784},$^1$ 
Jean Somalwar\orcidlink{0000-0001-8426-5732},$^1$
\newauthor
and Shriharsh P.~Tendulkar\orcidlink{0000-0003-2548-2926}$^{6,7}$ \\
$^1$Cahill Center for Astronomy and Astrophysics, MC\,249-17 California Institute of Technology, Pasadena CA 91125, USA. \\
$^2$ Department of Physics and Astronomy, West Virginia University, P.O. Box 6315, Morgantown, \\ WV 26506, USA \\
$^3$ Center for Gravitational Waves and Cosmology, West Virginia University, Chestnut Ridge Research Building, Morgantown, \\ WV 26505, USA \\
$^4$ Canadian Institute for Advanced Research, CIFAR Azrieli Global Scholar, MaRS Centre West Tower, \\ 661 University Ave. Suite 505, Toronto ON M5G 1M1, Canada \\
$^5$ Jet Propulsion Laboratory, California Institute of Technology, 4800 Oak Grove Dr, Pasadena, CA 91109, USA \\
$^6$ Department of Astronomy and Astrophysics, Tata Institute of Fundamental Research, Homi Bhabha Road, Colaba, \\ Mumbai, Maharashtra, 400005, India \\
$^7$ National Centre for Radio Astrophysics, Pune University Campus, Post Bag 3, Ganeshkhind, Pune, Maharashtra, 411007, India
}
\date{Accepted XXX. Received YYY; in original form ZZZ}

\pubyear{2021}

\begin{document}
\label{firstpage}
\pagerange{\pageref{firstpage}--\pageref{lastpage}}
\maketitle

\begin{abstract}

The physical properties of fast radio burst (FRB) host galaxies provide important clues towards the nature of FRB sources. The 16 FRB hosts identified thus far span three orders of magnitude in mass and specific star-formation rate, implicating a ubiquitously occurring progenitor object. FRBs localised with $\sim$arcsecond accuracy also enable effective searches for associated multi-wavelength and multi-timescale counterparts, such as the persistent radio source associated with FRB\,20121102A. 
Here we present a localisation of the repeating source \frb{}, and its association with a host galaxy (SDSS\,J050803.48$+$260338.0, $z=0.098$) and persistent radio source.
The galaxy is massive ($\sim3\times10^{10}\,M_{\odot}$), star-forming (few solar masses per year), and dusty. Very Large Array and Very Long Baseline Array observations of the persistent radio source measure a luminosity of $1.2\times10^{29}$\,erg\,s$^{-1}$\,Hz$^{-1}$, and show that is extended on scales $\gtrsim50$\,mas. We associate this radio emission with the ongoing star-formation activity in SDSS\,J050803.48$+$260338.0. Deeper, more detailed observations are required to better utilise the milliarcsecond-scale localisation of \frb{} reported from the European VLBI Network, and determine the origin of the large dispersion measure ($150-220$\,pc\,cm$^{-3}$) contributed by the host. SDSS\,J050803.48$+$260338.0 is an order of magnitude more massive than any galaxy or stellar system previously associated with a repeating FRB source, but is comparable to the hosts of so far non-repeating FRBs, further building the link between the two apparent populations.   

\end{abstract}














\section{Introduction}

The handful of fast radio bursts (FRBs) localised to individual galaxies comprises emerging evidence for a diversity of progenitor environments. The repeating source FRB\,20121102A is associated with a rapidly star-forming (0.13--0.23\,$M_{\odot}$\,yr$^{-1}$) dwarf (stellar mass of $\sim10^{8}M_{\odot}$) galaxy \citep{tbc+17,bta+17}, and two other repeating FRBs are also associated with galaxies that form stars $>10\times$ more rapidly than the Milky Way \citep{mnh+20,hps+20}. FRB\,20121102A is additionally associated with a luminous persistent radio source \citep{clw+17,mph+17} of unknown origin. Although the repeating source FRB\,20200120E has a comparable typical burst luminosity, repetition rate, and spectro-temporal characteristics to the repeating-FRB population \citep{bgk+21}, it is associated with an otherwise unremarkable globular cluster in the nearby galaxy M81 \citep{kmn+21}. FRB sources from which repetition has not been observed are located in galaxies with stellar masses ranging between $10^{9}-10^{11}\,M_{\odot}$, and star-formation rates up to a few $M_{\odot}$\,yr$^{-1}$ \citep{hps+20}. The FRB locations within the hosts are unremarkable \citep{mfs+20}, although a selection of biases remain to be untangled in their interpretation \citep{brd21}. 

The origins, life cycles, and ultimate fates of FRB progenitors remain unknown. The magnetospheres and immediate environments of neutron stars are the leading candidate FRB emission regions \citep[e.g.,][]{lkz20,mms+20}. An FRB-like burst has been associated with an active Galactic magnetar \citep{abb+20,brb+20}. However, observed phenomena like a several-day FRB periodic activity cycle \citep{aab+20}, the FRB source associated with a globular cluster \citep{kmn+21}, and the $\sim10^{29}$\,erg\,s$^{-1}$ compact persistent radio source (PRS) associated with FRB\,20121102A \citep{clw+17,mph+17} are difficult to reconcile with the known Galactic magnetar population \citep{kb17}. Attempts to jointly model FRB power sources and the emission mechanism typically invoke young magnetars with millisecond spin periods \citep[e.g.,][]{lk18,mms19,lbb20,l20}, or accretion from or interaction with a companion \citep[e.g.,][]{z18,smb+21}. The means by which such extreme systems form, and the lifetimes during which they produce the prodigious observed FRB rate \citep{r19}, continue to remain mysterious. Only through the continued characterisation of FRB host galaxies, and the positive or negative identification of multi-wavelength and multi-timescale counterparts, will this problem be addressed. 

The repeating source \frb\ was reported to be in an active state by \citet{atel14497} on 2021 March 21, with six bursts detected within five days. The dispersion measure (DM) of the source is 413.52$\pm$0.05\,pc\,cm$^{-3}$, with a Galactic-disk contribution of between 76\,pc\,cm$^{-3}$ \citep{cl02} and 109\,pc\,cm$^{-3}$ \citep{ymw17}. Over the next few months, further repeat bursts were detected at frequencies between 700\,MHz and 2\,GHz by the Australian Square Kilometre Array Pathfinder \citep[ASKAP;][]{askap1,askap2}, the Five hundred metre Aperture Spherical Telescope \citep[FAST;][]{fast1}, the Karl G. Jansky Very Large Array \textit{realfast} instrument \citep[VLA/\textit{realfast};][]{vla1}, the upgraded Giant Metrewave Radio Telescope \citep[uGMRT;][]{gmrt1,gmrt2}, the Stockert Radio Telescope \citep{stockert1}, the Onsala Radio Telescope \citep{onsala1}, and the Allen Telescope Array \citep{ata1}. Initial interferometric localisations with few-arcsecond accuracy were obtained within days to weeks of the CHIME/FRB announcement by the teams at ASKAP \citep{askap3,askap4}, VLA/\textit{realfast} \citep{vla1}, and uGMRT \citep{gmrt2}. A candidate host galaxy at a redshift $z=0.098\pm0.002$ \citep{askap5}, SDSS\,J050803.48$+$260338.0, was associated with \frb\ by \citet{askap3}. A milliarcsecond-accuracy localisation of \frb\ by the European VLBI Network \citep[EVN;][]{mkh+21}, reported on 2021 May 5, established the association beyond doubt.

Reports of a PRS associated with \frb\ based on 650\,MHz uGMRT observations \citep{gmrt1}, and 3\,GHz and 9\,GHz observations at the VLA \citep{rpp+21}, sparked particular excitement given the heretofore fruitless hunt for FRB/PRS associations since the case of FRB\,20121102A. The inferred radio luminosity of $\sim10^{29}$\,erg\,s$^{-1}$\,Hz$^{-1}$ at 1.4\,GHz is consistent with both the low end of the local luminosity function of radio AGN, and with typical galaxies in the local star-forming sequence \citep{ms07}. An absence of milliarcsecond-scale radio emission from the PRS was reported by \citet{mkh+21}, suggesting an extended emission region. 

In this paper, we present a comprehensive radio and optical study of \frb\ and its host galaxy. We begin in Section~\ref{obs} with a summary of observations by our group, including with the VLA, the Very Long Baseline Array (VLBA), and the Palomar 200-inch Hale Telescope (P200). We then extract critical parameters of the host galaxy and interpret the nature of the associated PRS in Section~\ref{interpretation}. A discussion of the implications for the source of \frb\ is presented in Section~\ref{discussion}, and we conclude in Section~\ref{conclusions}. Throughout, we adopt cosmological parameters from \citet{planck2018}, including a Hubble constant of $H_{0}=67.4$\,km\,s$^{-1}$\,Mpc$^{-1}$, a matter density parameter of $\Omega_{m}=0.315$, and a dark energy density parameter of $\Omega_{\Lambda}=0.685$.

\section{Observations}
\label{obs}

\subsection{VLA/\textit{realfast}}

We used the VLA (program code 21A-387) to observe \frb\ soon after \citet{atel14497} reported activity in the source. The FRB field was observed in ten, 52\,min sessions from 2021 April 05--15. The first two observing blocks were in L band (1--2\,GHz) and all others were observed in C band (4--8\,GHz). The VLA antennas were arranged in the D configuration, which has baseline lengths up to 1\,km.

We recorded visibility data with a sampling time of 5\,s while commensally streaming data with 10\,ms sampling time into the \textit{realfast} transient search system \citep{lbb+18}. We used \textit{realfast} to search for FRBs in real time with a typical 1$\sigma$\ sensitivity of 5\,mJy in 10\,ms. The search resampled the data to temporal widths of 10, 20, 40, and 80 ms in both bands and included DM trials up to 1500 and 3000\,pc\,cm$^{-3}$ at L and C band, respectively.

The \textit{realfast} system detected one burst in the first observing session on 2021 April 06 (burst MJD 59311.0129359, topocentric at 2.0\,GHz). The burst had a signal to noise ratio (S/N) of 26, a DM of $420\pm10$\,pc\,cm$^{-3}$, and was unresolved within the 10\,ms integration time. This significance was measured in a band from 1.3--1.5\,GHz that included all of the burst emission. We calibrated the data with the VLA calibration pipeline (version 2020.1) using flux calibrator 3C\,147. After applying these solutions, we measure a burst fluence of $2.4\pm0.1$\,Jy\,ms and position (J2000 epoch) ${\rm RA} = 05^{\rm h}08^{\rm m}03^{\rm s}.50$, ${\rm Dec} = 26^{\rm \circ}03'37''.71$. The source size is similar to the synthesised beam size of 55$''\times50''$\ at a position angle of $-2^{\rm \circ}$\ at 1.4\,GHz. The statistical position uncertainty is $0.8''$, while the total error is $1.9''$ (systematics dominated, see below).

The standard (slow) visibility data were imaged to search for persistent emission at the FRB location and estimate systematic position errors. We imaged a single observing session of 31\,min at L and C band with usable bandwidths of approximately 400 and 1500\,MHz, respectively.
Systematic source position error is typically dominated by antenna phase calibration errors that shift the centroid for all sources in the image. We estimate this effect by cross matching sources in the deep radio image to the PanSTARRS1 (PS1) catalogue \citep{cmm+16}. At L band, the confusion limited image does not have enough sources to perform a useful cross match. At C band, we find 5 cross matched sources with $2''$\ and estimate a systematic error of $0.5''$. Assuming this error scales with synthesised beam size, we estimate an L band systematic error of $1.7''$.

At L band, the naturally-weighted image shows no source brighter than 500\,$\mu$Jy at the FRB location. The noise level of the image is roughly 170\,$\mu$Jy, which is consistent with the expected confusion limit of this resolution at 1.4\,GHz \citep{ccf+12}. The C band image has a compact radio source with a peak flux density of $221\pm15\mu$Jy at position (J2000 epoch) ${\rm RA} = 05^{\rm h}08^{\rm m}03^{\rm s}.45$, ${\rm Dec} = 26^{\rm \circ}03'38''.00$. The statistical uncertainty in the position is $0.2''$, but the total error is dominated by systematic effects and is $0.5''$. The source size is $11.6''\times10.8''$\ at a position angle of $0^{\rm \circ}$\ at 5\,GHz, which is consistent with the synthesised beam size. Hereafter we denote this source PRS\,201124. 

\begin{figure}
    \centering
    \includegraphics[width=0.48\textwidth]{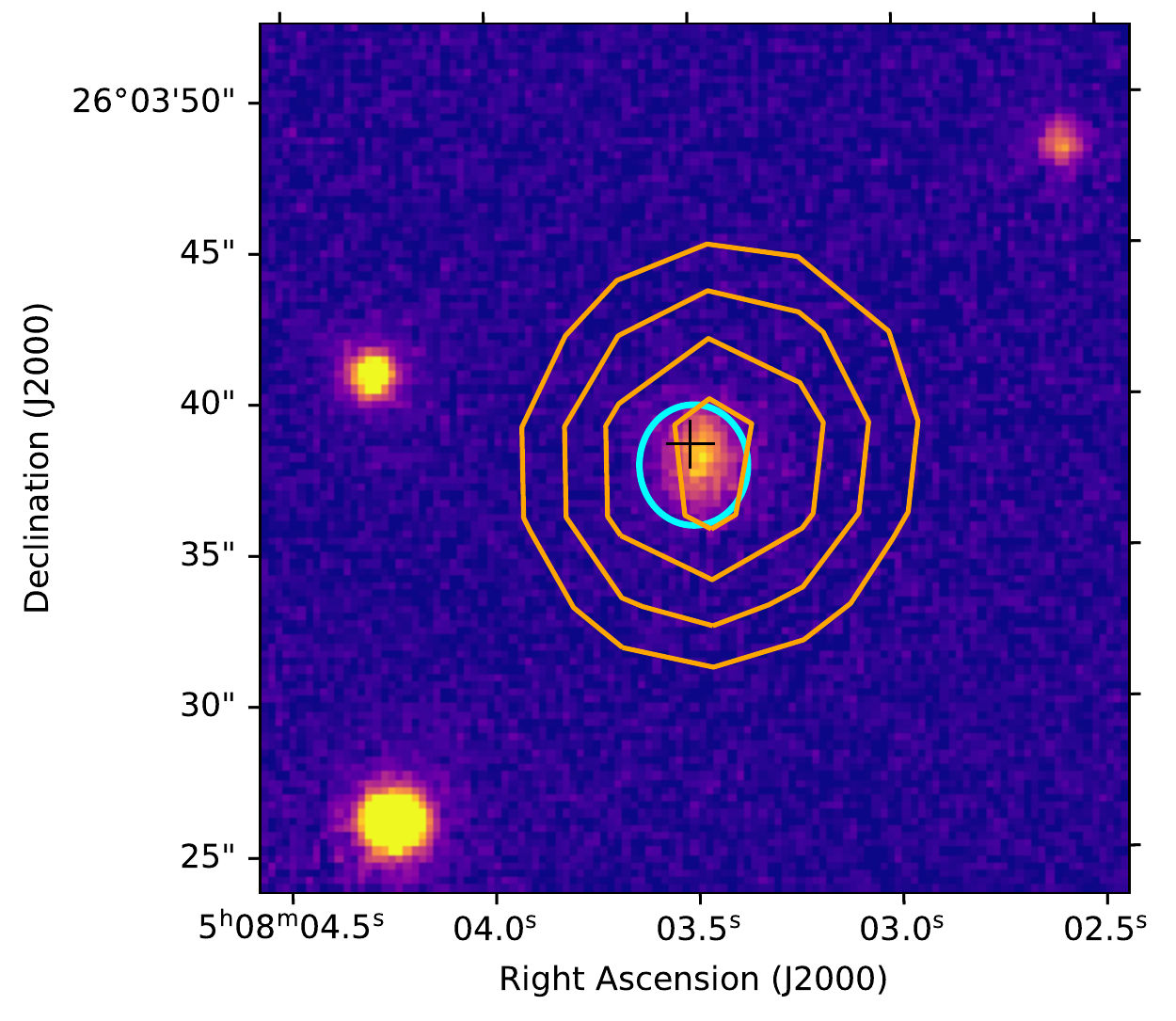}
    \caption{Localisation of \frb. \textit{Background:} stacked $i$-band image from the PanSTARRS Data Release 1 \citep{cmm+16}. \textit{Orange contours:} VLA C band image of the field, with contour levels of $6\sigma$, 9$\sigma$, $12\sigma$ and $15\sigma$, where $\sigma=13$\,$\mu$Jy/beam is the image noise level. \textit{Blue ellipse:} VLA/\textit{realfast} localisation error region of a burst from \frb. The dimensions represent the $1\sigma$ uncertainties in right ascension and declination. \textit{Black cross:} representation of the EVN position of bursts from \frb\ \citep{mkh+21}. The position error is much smaller than the arms of the cross.}
    \label{fig:overlay}
\end{figure}

Figure \ref{fig:overlay} shows the persistent L band burst location and C band emission overlaid on a PS1 image of the region. The burst and persistent emission are coincident with each other and the galaxy SDSS\,J050803.48$+$260338.0 (hereafter J0508$+$2603). No other optical source is consistent with the radio source locations. We use the burst localisation precision and PS1 catalogue to calculate the chance of association using the \texttt{astropath} Bayesian inference framework \citep{2021ApJ...911...95A}. Under a wide range of priors for host brightness and offset distribution, we find an association probability greater than 95\%. Consistent with previous analyses \citep{askap3,mkh+21}, we conclude that J0508$+$2603 is the host galaxy of \frb.

\subsection{VLBA L-band continuum}

We observed the location of \frb\ with the VLBA for 6\,hr between 2021 April 08 20:00 and 2021 April 09 02:00 UT. Data were recorded using the Digital Downconverter personalities of the ROACH Digital Backends with all ten antennas in four 64\,MHz wide spectral windows centred on 1407.75\,MHz, 1471.75\,MHz, 1663.75\,MHz, and 1727.75\,MHz, each with 128 channels. Data were recorded in dual circular polarisations with a data rate of 2.048\,Gbps at each station. The observations were phase referenced, with 71 scans of 210\,s duration on the target interspersed with 45\,s scans on the phase reference source J0500$+$2651 1.88$^{\circ}$ distant, for which we adopt a position \citep[International Celestial Reference Frame 3;][]{cjg+20} of ${\rm RA}=05^{\rm h}00^{\rm m}27^{\rm s}.87019156$, ${\rm Dec} = +26^{\circ}51'34''.3393223$ (J2000). Observations of the bandpass calibrator 3C\,84 (two 3\,min scans) and a check source J0502$+$2516 (two 45\,s scans) were also performed. The data were correlated at the NRAO Array Operations Center with the DiFX software correlator \citep{dbp+11}, with a target phase centre of ${\rm RA}=05^{\rm h}08^{\rm m}03^{\rm s}.50$, ${\rm Dec} = +26^{\circ}03'37''.8$ (J2000). 

We analysed the data using both CASA (version 5.6.1-8) and AIPS \citep{g03}, and obtained similar results. After data editing to excise radio-frequency interference, we derived initial bandpass solutions using 3C\,84, and complex gain solutions using global fringe fitting together with a single round of self-calibration on J0500$+$2651. We then applied the solutions to data on the target and the check source J0502$+$2516. Imaging was performed only in CASA using the \texttt{tclean} task, with the \citet{h74} deconvolver and robust weighting with a robustness parameter of 0.5. For the target, we present results from images made with a phase centre corresponding to the EVN position of \frb\ \citep{mkh+21}, over a $1.024\times1.024$\arcsec region. 

\begin{figure*}
    \centering
    \includegraphics[width=0.9\textwidth]{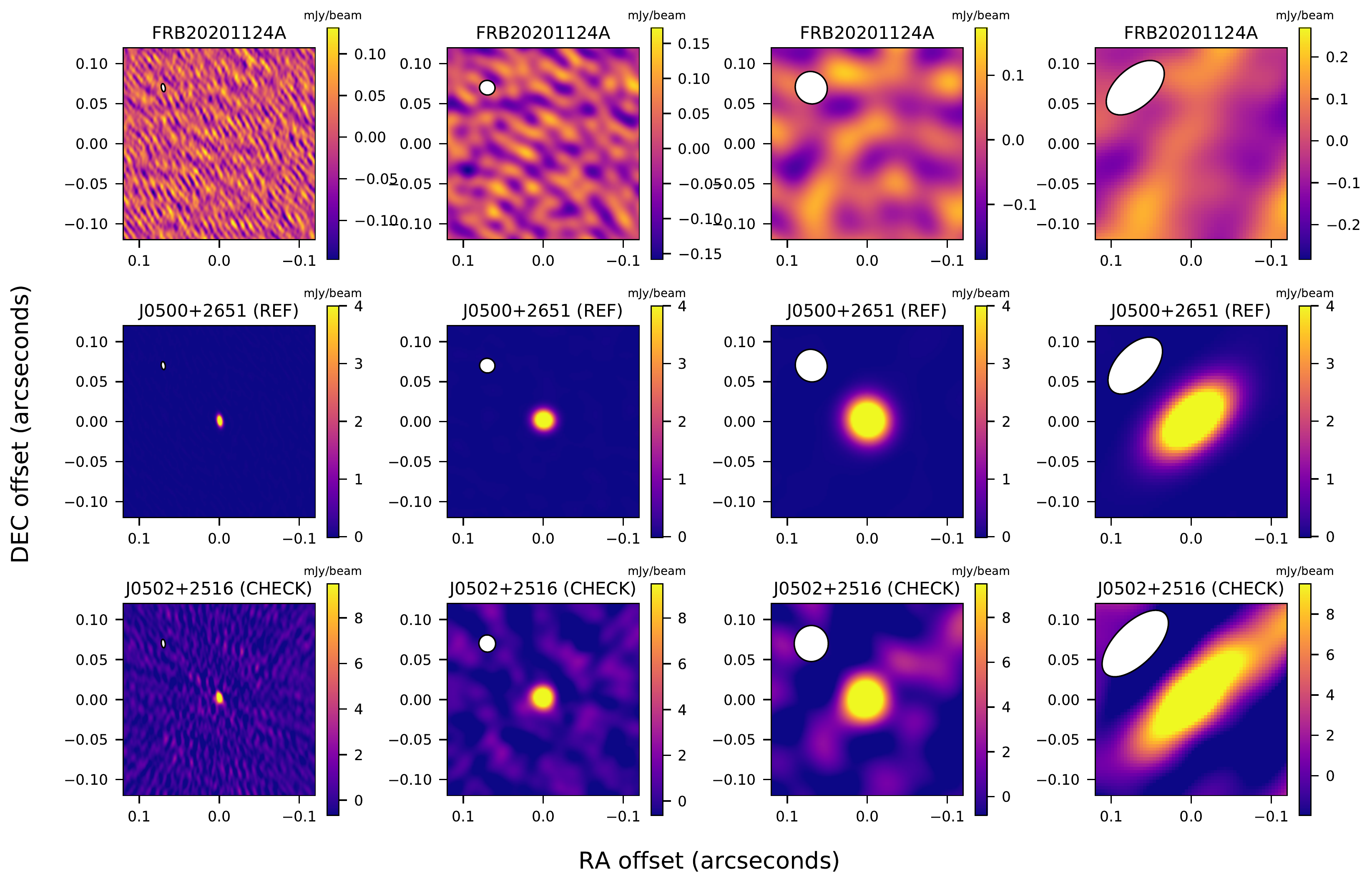}
    \caption{VLBA images of \frb\ (top row) at different angular scales, with images of the phase-reference (middle row) and independent check (bottom row) sources shown for comparison. From left to right, images were made with no $uv$ taper, and $uv$ tapers of 5000\,klambda, 2000\,klambda, and 1200\,klambda respectively (see Table~\ref{tab:1} for more details). Note that only two 45\,s scans were obtained on check source.}
    \label{fig:vlba}
\end{figure*}

No persistent emission was detected with the VLBA in any images made towards \frb. Further, no emission was detected towards the centre of light of J0508$+$2603. This result is consistent with the previous EVN observations \citep{mkh+21}, which concluded that PRS\,201124 is extended beyond milliarcsecond scales. We made a series of images with different Gaussian tapers in the $uv$ plane, as summarised in Table~\ref{tab:1}. The listed tapers correspond to the \texttt{tclean} `uvtaper' parameter, and we also list the full-width half-maximum (FWHM) of the synthesised beam and image noise rms in each case. The uvtaper parameter controls the width of a multiplicative Gaussian taper applied to the gridded visibilities to downweight data on longer baselines. In all cases, the minimum projected baseline length was 210\,km, corresponding to a largest angular scale of $\sim180$\,mas. In Figure~\ref{fig:vlba}, we show images of the target, phase reference, and check source made with each of the $uv$ tapers; note that all images were made with exactly the same calibrations applied to the data. 

\begin{table}
	\centering
	\caption{Summary of VLBA imaging results on \frb.}
	\label{tab:1}
	\begin{tabular}{ccc} 
		\hline
		$uv$ taper (k$\lambda$) & Beam FWHM (mas) & RMS (mJy\,beam$^{-1}$) \\
		\hline
		None & $10.2\times4.9$ & 0.04 \\
		5000 & $19.5\times18.2$ & 0.048 \\
		2000 & $41.4\times39.2$ & 0.062 \\
		1200 & $86.7\times48.7$ & 0.116 \\
		\hline
	\end{tabular}
\end{table}

In summary, we place $3\sigma$ upper limits on a compact radio source associated with \frb\ of between 0.12\,mJy and 0.2\,mJy on angular scales of between 5--50\,mas (see Table~\ref{tab:1}) for more detail). These flux-density limits correspond to radio luminosities of $3-9\times10^{28}$\,erg\,s$^{-1}$\,Hz$^{-1}$ at the distance of the \frb\ host J0508$+$2603.

\subsection{Palomar P200/DBSP}
\label{dbsp}

We observed the host galaxy of \frb, J0508$+$2603, with the P200 Double Spectrograph \citep[DBSP;][]{og82} on 2021 April 10 UT. Conditions were clear, with 1.2\arcsec~seeing. We obtained a low-resolution ($R\sim1000$) spectrum of J0508$+$2603 using the 600/4000 grating on the blue arm (central wavelength 4400\,\AA), and the 316/7500 grating on the red arm (central wavelength 7500\,AA). A 1.5\arcsec~slit was placed on the target, centred on the position ${\rm RA}=05^{\rm h}08^{\rm m}03^{\rm s}.48$, ${\rm Dec} = +26^{\circ}03'38''.0$ (J2000) at a position angle of 304.9$^{\circ}$. Two 1200\,s exposures were obtained with the blue arm, and three 800\,s exposures were obtained with the red arm, at a mean airmass of 1.5. The data were bias-subtracted, flat-fielded, cleaned of cosmic rays, wavelength calibrated using comparison-lamp spectra, sky-line subtracted, and optimally extracted using standard techniques implemented in a custom DBSP pipeline.\footnote{\url{https://github.com/finagle29/DBSP_DRP}} Flux calibration and telluric-line correction was performed with observations of the standard star Feige\,34 at a comparable airmass. 

\begin{figure*}
    \centering
    \includegraphics[width=0.8\textwidth]{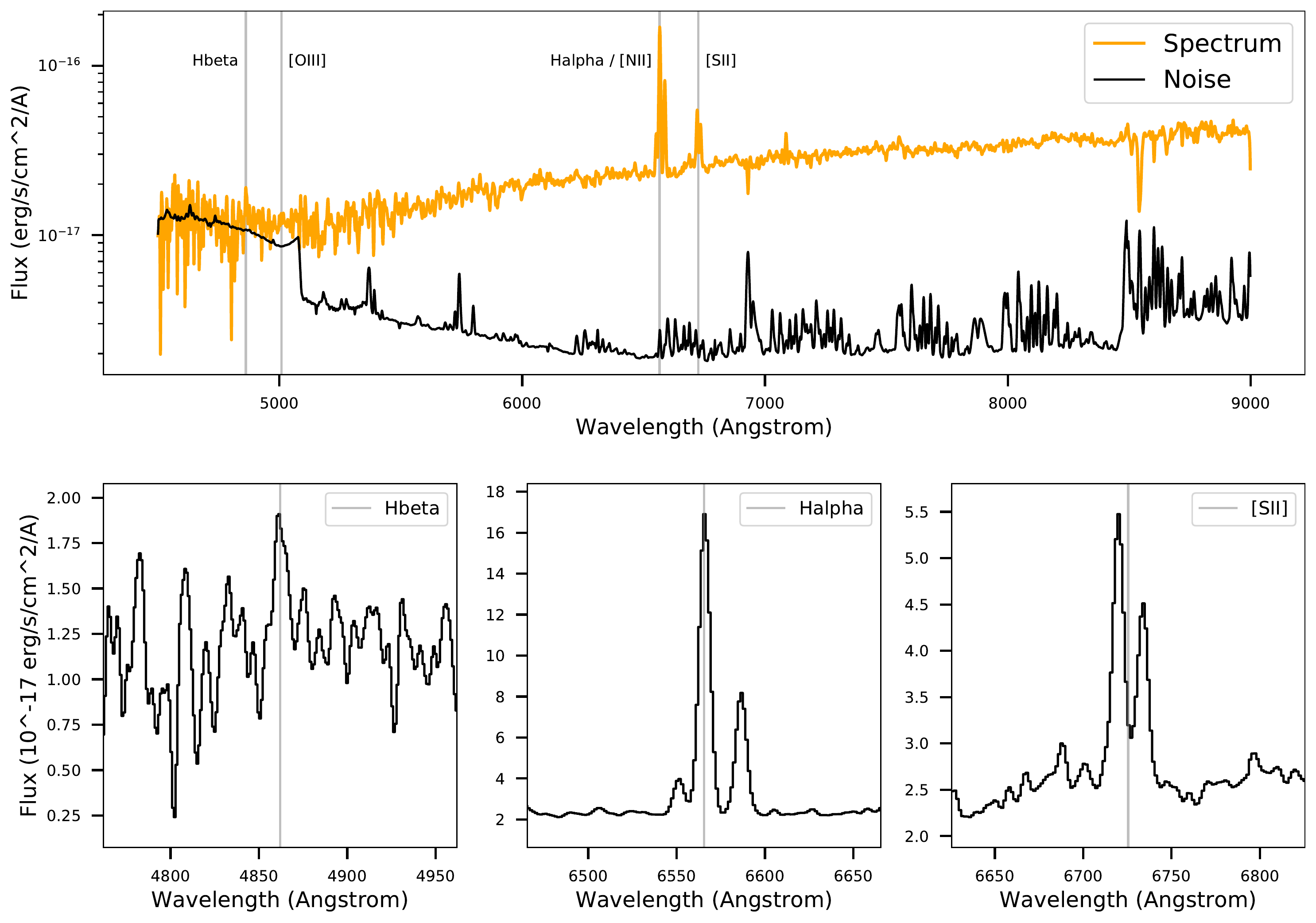}
    \caption{Palomar P200/DBSP spectrum of SDSS\,J050803.48$+$260338.0 (J0508$+$2603). \textit{Top:} flux calibrated spectrum (gold) and $1\sigma$ noise spectrum (black), with the locations of a few important spectral lines shown. \textit{Bottom, from left to right:} spectra centred on the H${\rm \beta}$, H${\rm \alpha}$ and [SII]\,$\lambda6718,6733$ lines. For better visualisation, all spectra are smoothed by a Gaussian with a standard deviation of 1.5 pixels. No extinction correction is applied to the data in this figure.}
    \label{fig:spectrum}
\end{figure*}

The spectrum of J0508$+$2603, shown in Figure~\ref{fig:spectrum}, exhibits a few clear emission-line features, including the ${\rm H\alpha}$ / [NII] complex, and the [SII]\,$\lambda6718,6733$ doublet. A tentative indication of ${\rm H\beta}$ is evident, although we do not claim a detection. After normalising by the continuum and correcting for Galactic extinction of $A_{V}=2.024$ \citep{sf11} assuming a \citet{fm07} extinction law, we measure line ratios of $\log ([{\rm NII}]/{\rm H\alpha}) = -0.40\pm0.01$ and $\log ([{\rm SII}]/{\rm H\alpha}) = -0.48\pm0.02$. Based on non-detections of the ${\rm H\beta}$ and [OIII]\,$\lambda5007$ lines,  we  derive a 95\% confidence upper limit on their ratio of $\log ([{\rm OIII}]/{\rm H\beta}) < 0.6$. According to the diagnostics of \citet{kgk+06}, these ratios are consistent with a softer source of ionising radiation corresponding to young stars, i.e., corresponding to ongoing star formation. 

We caution that the above analysis of the P200/DBSP observations is not fully representative of the J0508$+$2603. The spectrum does not capture all the light from the galaxy, and we have applied no corrections for slit losses. We have also not attempted to model absorption features at the ${\rm H\alpha}$ and ${\rm H\beta}$ wavelengths corresponding to stellar photospheres. Finally, we do not spatially resolve the galaxy in the spectroscopic observations.

\section{Interpretation}
\label{interpretation}

\subsection{The optical/infrared SED and spectrum of J0508+2603}

\begin{figure*}
    \centering
    \includegraphics[width=0.8\textwidth]{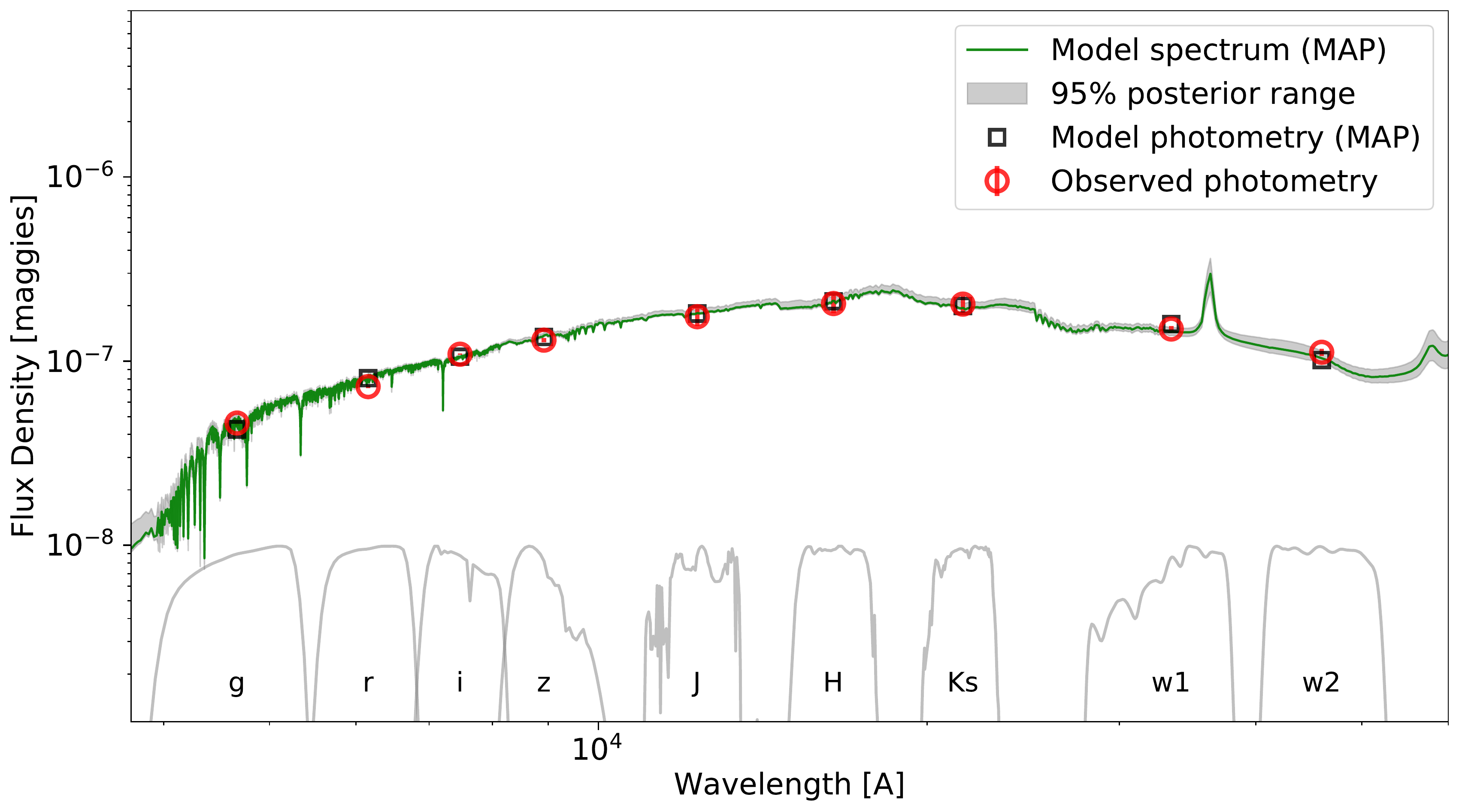}
    \caption{Observed and modelled spectral energy distribution (SED) of SDSS\,J050803.48$+$260338.0 (J0508$+$2603). Observed flux densities in various filters (representative transmission curves shown in grey), with $1\sigma$ error bars, are shown as red circles. The maximum aposteriori probability (MAP) model SED derived from \texttt{Prospector} is shown as a green curve, and the corresponding model photometry in all filters is shown as black squares. The grey shading indicates the 95\% confidence range in the modelled SED at every wavelength. Note that no nebular emission was included in the model. All points and curves are corrected for extinction in the Milky Way interstellar medium.}
    \label{fig:sed}
\end{figure*}

We collated archival photometric observations of J0508$+$2603 to perform stellar population synthesis modelling of its spectral energy distribution (SED). J0508$+$2603 appears in all bands in the Sloan Digital Sky Survey (SDSS) Data Release 16 \citep[DR16;][]{apa+20}, the Two Micron All Sky Survey \citep[2MASS;][]{scs+06}, and in the ALLWISE data release \citep{cwc+21} from the Wide-field Infrared Survey Explorer (WISE). These data, together with results from the modelling described below, are shown in Figure~\ref{fig:sed}. We discarded the $u$-band SDSS DR16 catalogue measurement of $23.09\pm0.55$ magnitudes, as no source was evident in a visual inspection of the image. We also did not attempt to model thermal emission in the WISE $w3$ and $w4$ bands, given that model uncertainties in these emission bands are typically large and hard to quantify \citep[e.g.,][]{ljc+17,brd+20}.

We used the \texttt{Prospector} \citep{ljc+17,jlc+21} stellar population inference code to model the SED of J0508$+$2603. \texttt{Prospector} enables efficient sampling of the posterior distributions of model parameters that describe the stellar populations of galaxies. The forward model is built on the Flexible Stellar Population Synthesis code \citep{cgw09}. We used a standard `delay-tau' parametric star-formation history, and sampled from the posterior using \texttt{emcee} \citep{2013PASP..125..306F}. A model for dust attenuation and re-radiation was also included. Priors used included a log-uniform prior on the mass in formed stars (hereafter the stellar mass) of between $10^{7}-10^{13}M_{\odot}$, a top-hat prior on the internal dust extinction ($A_{V}$) of $0-3$ magnitudes, a top-hat prior on the age of the stellar population of between $0.001-13.8$\,Gyr, a log-uniform prior on the star-formation timescale of between $0.001-30$\,Gyr, and a top-hat prior on the ratio of the metallicity to the solar metallicity ($\log Z_{\rm sol}$) of between $-2$ and $0.2$. We found  $\log(M/M_{\odot})=10.62^{+0.07}_{-0.06}$,  $\log Z_{\rm sol}=-0.9\pm0.2$, and $A_{V}=1.5\pm0.2$ magnitudes. This metallicity is remarkably low for such a massive galaxy, as evidenced by the position of this galaxy 0.8\,dex below the mass-metallicity relation in the local Universe \citep{cmc+20}. Furthermore, the large ratio of $\log ([{\rm NII}]/{\rm H\alpha})$ is consistent with approximately solar metallicity \citep{pp04}. We therefore re-ran the \texttt{Prospector} model with a Gaussian prior on $\log Z_{\rm sol}$ with mean $0.012$ and standard deviation $0.205$, based on the \citep{pp04} relation between $\log ([{\rm NII}]/{\rm H\alpha})$ and metallicity. This yielded $\log(M/M_{\odot})=10.48^{+0.03}_{-0.05}$, $\log Z_{\rm sol}=-0.3^{+0.2}_{-0.3}$ and $A_{V}=1.3\pm0.2$. We adopt these latter values as representative of J0508$+$2603. 

The star-formation history was less well constrained by the data, with highly correlated posterior distributions of the delay and timescale parameters. We explored non-parametric models for the star-formation history within \texttt{Prospector} \citep{lcj+19}, and found similar issues. However, it is clear that the galaxy is young, with a 90\% confidence upper limit on the delay parameter in the delay-tau model of 3\,Gyr; this is also evidenced by the low metallicity \citep{gcb+15}. The maximum aposteriori probability model, together with an indication of the range of possible models, is shown in Figure~\ref{fig:sed}. 

With the measurement of the internal dust extinction, an estimate of the star-formation rate (SFR) corresponding to the ${\rm H\alpha}$ luminosity is possible. Combining Galactic and internal extinction, and adopting a luminosity distance of 467.3\,Mpc, the P200/DBSP spectrum implies an ${\rm H\alpha}$ luminosity of $L_{\rm H\alpha}=(3.4\pm0.8)\times10^{41}$\,erg\,s$^{-1}$. Assuming a Balmer decrement of 3, the predicted ${\rm H\beta}$ flux is $4.3\times10^{-15}$\,erg\,s$^{-1}$\,cm$^{-2}$\,\AA$^{-1}$, which is just $\sim3\times$ higher than the noise level in the P200/DBSP spectrum. This is consistent with the observed spectrum (Figure~\ref{fig:spectrum}). For consistency with the work of \citet{hps+20}, we adopt their conversion between $L_{\rm H\alpha}$ and SFR to derive an SFR of $\sim1.7$\,$M_{\odot}$\,yr$^{-1}$, with an uncertainty of $\sim35\%$ (including uncertainty in the $L_{\rm H\alpha}$-SFR relation). The specific SFR of J0508$+$2603 is thus $\sim8\times$ that of the Milky Way. We emphasise that, as discussed in Section~\ref{dbsp}, our ${\rm H\alpha}$-based SFR is likely to be a lower limit. 

A picture thus emerges of J0508$+$2603 as a young galaxy just a factor of a few less massive than the Milky Way. A high specific SFR is observed, together with significant internal dust extinction that partially attenuates the observed star formation. The internal extinction is indeed larger than in 90\% of SDSS galaxies with similar ${\rm H\alpha}$ luminosities \citep{xww+12}. Better constraints on the metallicity would be derived with more complete line-ratio diagnostics from a deeper spectrum.

\subsection{PRS 201124 represents star-formation activity}

\begin{figure}
    \centering
    \includegraphics[width=0.48\textwidth]{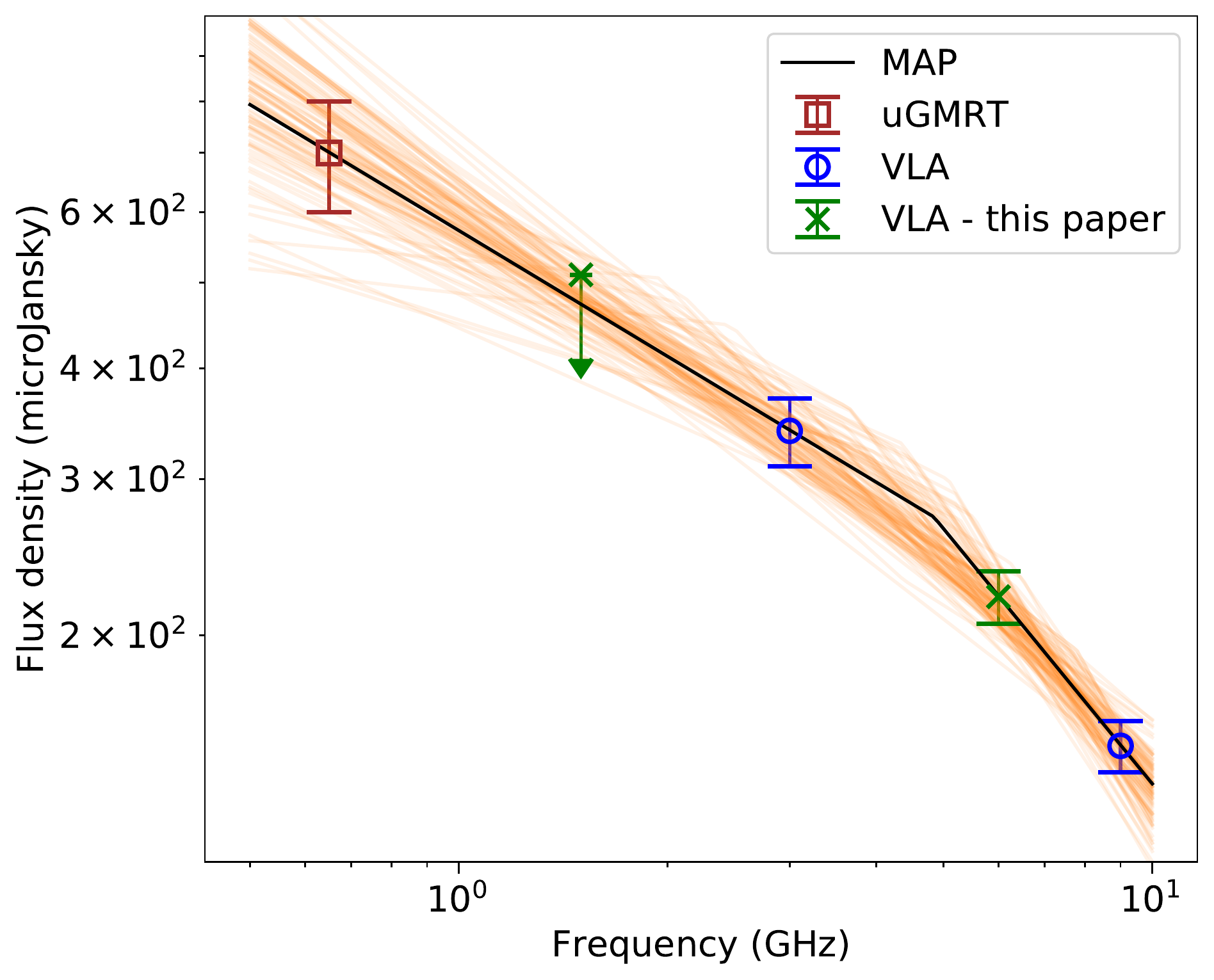}
    \caption{Radio spectrum of PRS\,201124. The points indicate measurements presented here (green cross), and additional measurements from the uGMRT \citep[brown square;][]{gmrt1} and VLA \citep[blue circles;][]{rpp+21}. We fit a four-parameter broken power-law model to the four measurements using the \texttt{emcee} package \citep{fhl+13}, and the maximum aposteriori probability (MAP) model is shown as a black line. Models corresponding to 100 random draws from the posterior are shown as orange curves.}
    \label{fig:radiospec}
\end{figure}

The SED of PRS\,201124, based on our VLA observations and those of \citet{rpp+21} and \citet{gmrt1}, is shown in Figure~\ref{fig:radiospec}. Our observations, together with those from the EVN \citep{mkh+21}, clearly demonstrate that PRS\,201124 is extended. For example, no emission is observed in our VLBA images on scales of $\lesssim50$\,mas (94\,pc at the distance of J0508$+$2603), with a $3\sigma$ upper limit of 0.2\,mJy or $5\times10^{28}$\,erg\,s$^{-1}$\,Hz$^{-1}$. Interpolating between the 650\,MHz and 3\,GHz observations of PRS\,201124 \citep{gmrt1,rpp+21}, the implied total flux density of PRS\,201124 at 1567.75\,MHz (the midpoint of the VLBA band) is $\sim0.46$\,mJy, or $1.2\times10^{29}$\,erg\,s$^{-1}$\,Hz$^{-1}$. For comparison, the only previously reported PRS associated with an FRB (FRB\,20121102A) had a 1.77\,GHz luminosity on milliarcsecond scales of $2\times10^{29}$\,erg\,s$^{-1}$\,Hz$^{-1}$ \citep{mph+17}. We conclude that a PRS like that associated with  FRB\,20121102A is not present at the location of \frb.

On the other hand, the radio luminosity of PRS\,201124 is consistent with the observed SFR of J0508$+$2603. The SFR derived from the ${\rm H\alpha}$ luminosity implies a 1.4\,GHz radio luminosity of $L_{\rm 1.4\,GHz}\sim3\times10^{28}$\,erg\,s$^{-1}$\,Hz$^{-1}$ from the relation of \citet{b03}. The relation of \citet{mcs+11}, which includes a factor that depends on the linear extent of the emission ($D$), implies $L_{\rm 1.4\,GHz}\sim2\times10^{28}[D/(3\,{\rm kpc})]^{0.75\pm0.11}$\,erg\,s$^{-1}$\,Hz$^{-1}$ with an uncertainty of $\sim1$\,dex. For the reasons discussed in Section~\ref{dbsp}, it is likely that the ${\rm H\alpha}$ luminosity of J0508$+$2603 underestimates the total SFR. Furthermore, radio-derived SFRs are averaged over $\sim100$\,Myr \citep{c92}, whereas the ${\rm H\alpha}$ luminosity represents more recent star formation. If the SFR is declining with time, the radio-derived SFR will exceed that derived from $L_{\rm H\alpha}$. It is therefore not surprising that the observed luminosity of PRS\,201124 mildly exceeds the predicted luminosity from the ${\rm H\alpha}$-derived SFR. The luminosity of PRS\,201124 may indeed represent a more accurate estimate of the SFR, i.e., of $\sim7$\,$M_{\odot}$\,yr$^{-1}$. 

The radio SED of PRS\,201124 is also consistent with typical star-forming radio galaxies \citep{klv18,tsd+19}. We model the SED (Figure~\ref{fig:radiospec}) using a broken power-law function \citep[Equation~(3) of][]{tsd+19}, with distinct spectral indices $\alpha$ and $\beta$ respectively below and above a break frequency $\nu_{b}$: 
\begin{equation}
    \log F_{\nu} = 
    \begin{cases}
    -\alpha\log(\nu/\nu_{b}) + F_{0}, & \text{if } \nu\leq \nu_{b}\\
    -\beta\log(\nu/\nu_{b}) + F_{0}, & \text{if } \nu> \nu_{b}.
    \end{cases}
\end{equation}
Here, $F_{\nu}$ is flux density, $\nu$ is frequency, and $F_{0}$ is a reference flux density. Broken power laws are expected in the case of significant cooling of the synchrotron-emitting electrons. We use \texttt{emcee} to fit this four-parameter model to the four data points, and find $\alpha=0.48^{+0.1}_{-0.18}$, $\beta=0.96^{+0.29}_{-0.16}$ and $\nu_{b}=5.0^{+2.2}_{-2.4}$\,GHz. Broken power-law spectra like this are observed in several nearby star-forming galaxies \citep{klv18}. Using a large sample of galaxies with SFRs at redshifts $0.3<z<4$ in excess of 100\,$M_{\odot}$\,yr$^{-1}$ from the VLA-COSMOS survey, \citet{tsd+19} find $\alpha=0.53\pm0.04$, $\beta=0.94\pm0.06$, and $\nu_{b}=4.3\pm0.6$\,GHz. Our measurements are in good agreement with this sample of highly star-forming galaxies. 

The extended morphology, luminosity and SED of PRS\,201124 all correspond to the star-formation activity inferred in J0508$+$2603. As discussed in Section~\ref{dbsp}, the emission-line ratios in our P200/DBSP spectrum of J0508$+$2603 show no evidence for an AGN, further establishing this correspondence. At the highest angular resolution of our VLBA observations (see Table~\ref{tab:1}), we place a $3\sigma$ limit on the 1567.75\,MHz luminosity of a compact PRS associated with \frb\ of $3\times10^{28}$\,erg\,s$^{-1}$\,Hz$^{-1}$, nearly an order of magnitude below the 1.77\,GHz luminosity of the PRS associated with FRB\,20121102A \citep{mph+17}. 

\section{Discussion}
\label{discussion}

\begin{figure}
    \centering
    \includegraphics[width=0.48\textwidth]{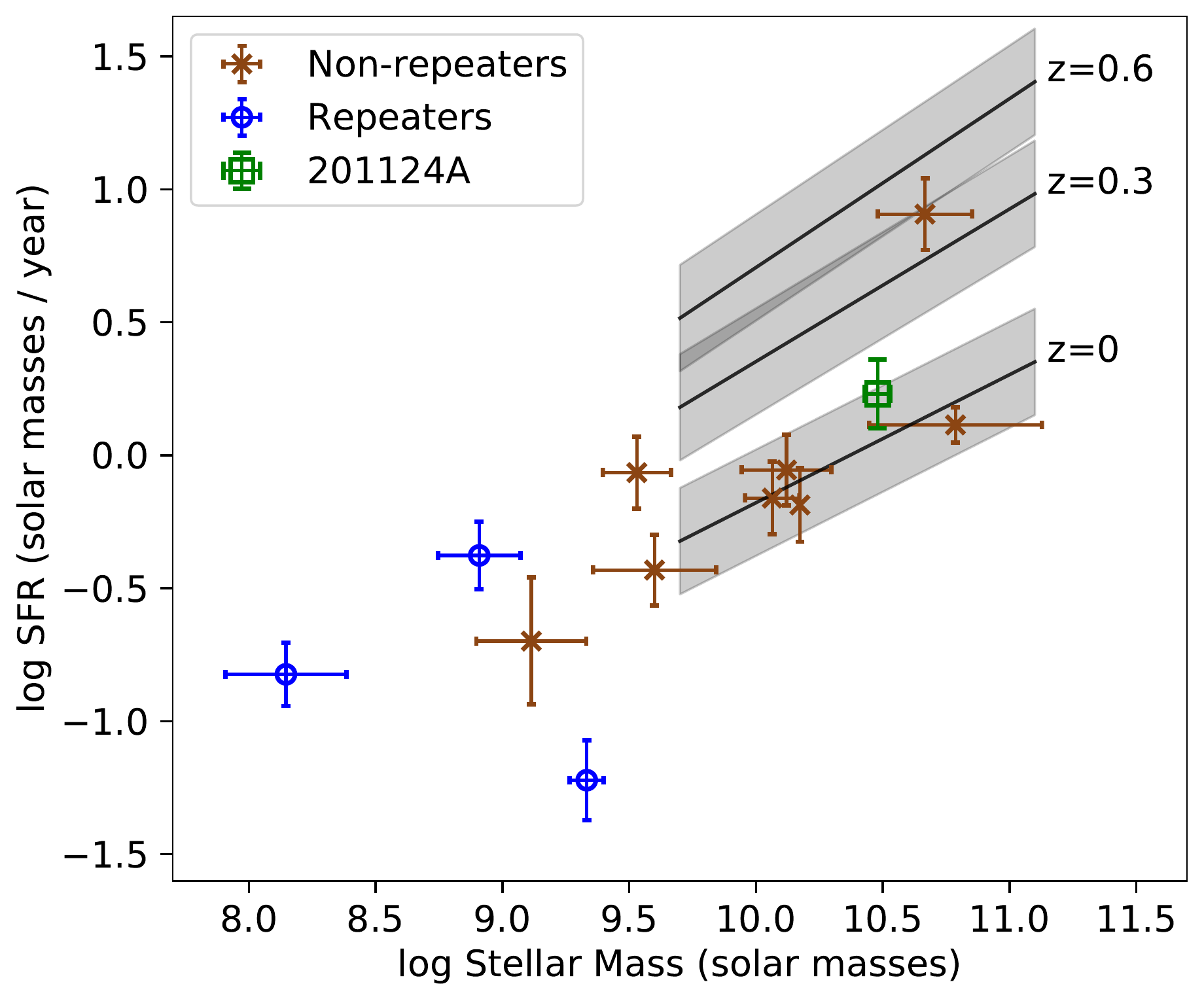}
    \caption{Stellar masses and SFRs of FRB host galaxies. The background sample of repeating and so far non-repeating FRB host data are from \citet{hps+20} (see text for details on excluded FRBs), with the exception of FRB\,190523 for which we use the [OII]-derived SFR \citep{rcd+19} of $1.3\pm0.2$\,$M_{\odot}$\,yr$^{-1}$. The three black lines indicate fits to the star forming main sequence of galaxies at different redshifts; the shaded regions indicate the $1\sigma$ intrinsic scatter \citep{ssc+14}.} 
    \label{fig:msfr}
\end{figure}

The host galaxy of \frb, J0508$+$2603, is unremarkable within the diverse range of FRB host galaxies \citep{hps+20}. Among the sample in hand, stellar masses range from $10^{8}-10^{11}\,M_{\odot}$, and star formation is evident in all but two cases \citep{rcd+19,kmn+21}. Specific SFRs range from an order of magnitude below that of the Milky Way, to three orders of magnitude above that of the Milky Way. Metallicities range from sub-solar in the case of FRB\,20121102A \citep[$\log Z_{\rm sol}<-0.58$;][]{tbc+17}, to approximately solar in the remainder of the sample in hand. J0508$+$2603 is among the more massive FRB host galaxies, and the most massive host of an FRB that is observed to repeat (Figure~\ref{fig:msfr}).\footnote{In making this statement, we do not consider M81 as the true host of FRB\,20200120E, but rather associate it directly with its host globular cluster \citep{kmn+21}}. We estimate a DM contributed by the host of $150-220$\,pc\,cm$^{-3}$, assuming a fiducial fraction of cosmic baryons in the intergalactic medium of 0.7 \citep{sd18}, a Milky Way halo contribution of between $50-80$\,pc\,cm$^{-3}$ \citep{pz19}, and a range of Milky Way disk DM contributions between the \citet{cl02} and \citet{ymw17} models. This is larger than is observed in most FRBs localised to host galaxies \citep{mpm+20}, and consistent with the host DM inferred for FRB\,20121102A \citep{tbc+17}. Further interpretation of this result will require high spatial resolution optical/infrared (OIR) observations of J0508$+$2603 to estimate the possible path length of the FRB through the host interstellar medium, which will enable any egregious local DM contributions to be identified.    

As just the fifth reported host system of a repeating FRB, our analysis of J0508$+$2603 further demonstrates the wide range of possible hosts of active FRB sources. Indeed, the J0508$+$2603 is at the more massive end of the star-forming main sequence of galaxies, consistent with the hosts of so far non-repeating FRBs but distinct from the remainder of the repeating-FRB host sample (Figure~\ref{fig:msfr}). This empirical fact further strengthens the link between the sources of actively repeating and so far non-repeating FRBs \citep{r19,jof+20}. We caution, however, against detailed statistical inference from the data in Figure~\ref{fig:msfr}. The star-formation rates and stellar masses are measured with different methods, with for example different amounts of the host-galaxy light included in the measurements of ${\rm H\alpha}$ luminosities. Sporadic attempts have been made to separate nuclear activity from star-formation activity. Some FRBs have been excluded from Figure~\ref{fig:msfr} because of insecure host associations \citep[FRB 190611;][]{hps+20}, incomplete data \citep[FRB\,190614;][]{lbp+20}, and low luminosity \citep[FRBs 200428 and 20200120E;][]{brb+20,kmn+21}. 

The stellar mass and SFR of J0508$+$2603 provide some insight into the source of \frb. We can compare these properties to the samples of core-collapse supernovae (CCSNe), superluminous supernovae (SLSNe), and long gamma-ray bursts (LGRBs) assembled by \citet{tp21}. These authors corrected the distributions of stellar mass and SFR for cosmic evolution, which is critical in comparing them to the local-Universe galaxy distribution \citep[see also][]{brd21}. Only two SLSNe out of 53 (one of each of types I and II) and no LGRBs (out of 17) within the \citet{tp21} sample are found in galaxies as massive as J0508$+$2603. On the other hand, the properties of J0508$+$2603 place it between the 80$^{\rm th}$ and 90$^{\rm th}$ percentiles of the stellar-mass and SFR distributions of the hosts of CCSNe. Formation channels for the \frb\ source associated with stellar evolutionary pathways associated with LGRBs and SLSNe are therefore not implicated by the properties of J0508$+$2603, consistent with previous findings based on the FRB host-galaxy population \citep{brd21}. 

We anticipate that more detailed observations of J0508$+$2603 will yield significant further insights. In particular, space-based OIR observations with high angular resolution will enable a direct comparison with observations of the hosts of other repeating FRBs. Specifically, in combination with the EVN localisation of \frb\ \citep{mkh+21}, this will enable the local environment of the FRB source to be placed in the context of the host-galaxy structure, and allow the host DM to be better interpreted \citep{2021ApJ...908L..12T}. A deeper, possibly spatially resolved OIR spectrum of J0508$+$2603 will enable more accurate metallicity and SFR measurements, in turn providing more robust measurements of the stellar mass and star-formation history. Finally, radio observations of J0508$+$2603 / PRS\,201124 with $0.1-1$\arcsec~resolution will likely resolve the morphology of the PRS, better establishing its nature and link with \frb. 

\section{Conclusions}
\label{conclusions}

We present a study of the host galaxy and persistent radio counterpart of the repeating \frb. We conclude the following:
\begin{enumerate}

    \item \frb\ is associated with a galaxy (J0508$+$2603) at $z=0.098$. J0508$+$2603 has a stellar mass of $\log(M/M_{\odot})=10.48^{+0.03}_{-0.05}$, a ratio between its metallicity and the solar metallicity of $\log Z_{\rm sol}=-0.3^{+0.2}_{-0.3}$, and a significant internal dust extinction of $A_{V}=1.3\pm0.2$. Based on an extinction-corrected ${\rm H\alpha}$ luminosity of $L_{\rm H\alpha}=(3.4\pm0.8)\times10^{41}$\,erg\,s$^{-1}$ (not corrected for slit losses or the presence of absorption in the stellar continuum), we derive a star-formation rate of $\sim1.7$\,$M_{\odot}$\,yr$^{-1}$ (35\% $1\sigma$ uncertainty).
    
    \item We find that the persistent radio source (PRS\,201124) associated with \frb\ is extended on scales $\gtrsim50$\,mas (94\,pc at the distance of J0508$+$2603). Considering the spatial extent, luminosity, and SED of PRS\,201124, the persistent emission is consistent with non-thermal emission caused by the observed ongoing star-formation activity. We place an upper bound on the luminosity of a compact ($\lesssim10$\,pc) PRS at the location of \frb\ of $3\times10^{28}$\,erg\,s$^{-1}$\,Hz$^{-1}$. Future searches for PRSs at the locations of FRBs should be careful to exclude radio sources originating in star-formation, in addition to nuclear activity. 
    
    \item The host galaxy of \frb, J0508$+$2603, is more massive (by an order of magnitude) than any previous known host galaxy of a repeating FRB, but has a comparable stellar mass and SFR to known hosts of so far non-repeating FRBs. This provides further evidence for commonality between the sources of repeating and so far non-repeating FRBs. 
    
    \item The stellar mass of J0508$+$2603 is much larger than the typical host galaxies of SLSNe and LGRBs, but together with the SFR is consistent with CCSNe host galaxies. 
    
\end{enumerate}
More detailed studies of J0508$+$2603 and PRS\,201124, with higher sensitivity and a wider range of angular resolutions in the radio and OIR bands, are required to better interpret the exquisite data in hand on \frb. Larger samples of localised FRBs with systematic host-galaxy studies will continue to refine our understanding of the origins of FRBs. 

\section*{Acknowledgements}

We thank the staff of the VLBA for rapidly scheduling and executing the observations presented here. This research was supported by the National Science Foundation under grant AST-1836018 and AST-2022546. K.A.~and S.B.S.~acknowledge support from NSF grant AAG-1714897. S.B.S~is a CIFAR Azrieli Global Scholar in the Gravity and the Extreme Universe program.
The NANOGrav project receives support from National Science Foundation (NSF) Physics Frontier Center award number 1430284.
Part of this research was carried out at the Jet Propulsion Laboratory, California Institute of Technology, under a contract with the National Aeronautics and Space Administration. The National Radio Astronomy Observatory is a facility of the National Science Foundation operated under cooperative agreement by Associated Universities, Inc.

\section*{Data Availability}

Original data presented herein include VLA, VLBA, and P200/DBSP observations. The VLA and VLBA primary data products are archived at the NRAO Data Archive (\url{https://archive.nrao.edu/archive/advquery.jsp}). The VLA secondary (\textit{realfast}) data products are will soon be available through the NRAO archive, but are currently available on request. The raw data obtained during the P200/DBSP observations are available upon request. All data that shown in figures are available upon request. 



\bibliographystyle{mnras}
\bibliography{frb} 





\bsp	
\label{lastpage}
\end{document}